\documentclass[letterpaper]{article} % DO NOT CHANGE THIS
\usepackage[table, xcdraw]{xcolor}
\usepackage{aaai23}  % DO NOT CHANGE THIS
\usepackage{times}  % DO NOT CHANGE THIS
\usepackage{helvet}
\usepackage{courier}  % DO NOT CHANGE THIS
\usepackage[hyphens]{url}  % DO NOT CHANGE THIS
\usepackage{graphicx} % DO NOT CHANGE THIS
\usepackage{natbib}  % DO NOT CHANGE THIS AND DO NOT ADD ANY OPTIONS TO IT
\usepackage{caption} % DO NOT CHANGE THIS AND DO NOT ADD ANY OPTIONS TO IT
\usepackage{algorithm}
\usepackage[noend]{algorithmic}
\usepackage{microtype}
\usepackage{enumitem}
\usepackage{newfloat}
\usepackage{listings}
\DeclareCaptionStyle{ruled}{labelfont=normalfont,labelsep=colon,strut=off} % DO NOT CHANGE THIS
\floatstyle{ruled}
\newfloat{listing}{tb}{lst}{}
\floatname{listing}{Listing}
\lstdefinestyle{dsl}
{
	language=csh,
	linewidth=\columnwidth,
	stringstyle={\color{gray}},
	escapeinside={(*}{*)},
	basicstyle=\footnotesize\ttfamily,
	keywordstyle=\bfseries,
	morekeywords={@start,@output,@input,string,bool,string[],int,Tuple,<,>,language,@values,using,Regex},
	morecomment=[l][\color{gray}]{//}
}

\usepackage{amsmath}
\usepackage{amsfonts}
\usepackage{booktabs}
\usepackage{tabularx}

\usepackage{tikz}

\newcommand{\system}[0]{\textsc{ring}}
\definecolor{babypink}{rgb}{0.96, 0.76, 0.76}
\definecolor{celadon}{rgb}{0.67, 0.88, 0.69}
\definecolor{cinnabar}{rgb}{0.89, 0.26, 0.2}

\usepackage{tikz}
\usepackage{tikzscale}
\usepackage{pgfplots}
\pgfplotsset{width=10cm,compat=1.9}

\usepackage{multirow}
\usepackage{booktabs}
\usepackage{nicematrix}
\usepackage{enumitem}
\usepackage{amsmath,amssymb}

\usepackage{subcaption}

\DeclareCaptionStyle{ruled}{labelfont=normalfont,labelsep=colon,strut=off} %
\floatstyle{ruled}
\newfloat{listing}{tb}{lst}{}
\floatname{listing}{Listing}

\usepackage{tcolorbox}
\newcommand{\lcolorbox}[2]{%
	\hspace*{-\fboxsep}\colorbox{#1}{#2}%
}

\title{Repair Is Nearly Generation: Multilingual Program Repair with LLMs}

\author {
    Harshit Joshi\textsuperscript{\rm 1},
    Jos\'e Cambronero\textsuperscript{\rm 2}\thanks{Listed in alphabetical order},
    Sumit Gulwani\textsuperscript{\rm 2}\footnotemark[1],
    Vu Le\textsuperscript{\rm 2}\footnotemark[1],
    Ivan Radi\v{c}ek\textsuperscript{\rm 3}\footnotemark[1],
    Gust Verbruggen\textsuperscript{\rm 4}\footnotemark[1]
}
\affiliations {
    \textsuperscript{\rm 1} Microsoft, India\\
    \textsuperscript{\rm 2} Microsoft, USA\\
    \textsuperscript{\rm 3} Microsoft, Croatia\\
    \textsuperscript{\rm 4} Microsoft, Belgium\\
    
    \{t-hjoshi, jcambronero, sumitg, levu, ivradice, gverbruggen\}@microsoft.com
}

\begin{document}

\maketitle

\begin{abstract}

Most programmers make mistakes when writing code. Some of these mistakes
are small and require few edits to the original program -- a class
of errors recently termed \emph{last mile mistakes}. These errors break
the flow for experienced developers and can stump novice programmers.
Existing automated repair techniques targeting this class of errors
are language-specific and do not easily carry over to new languages.
Transferring symbolic approaches requires
substantial engineering and neural approaches require data and retraining. We introduce
\system{}, a multilingual repair engine powered by a large language model
trained on code (LLMC) such as Codex. Such a multilingual engine enables a \emph{flipped model}
for programming assistance, one where the programmer writes code
and the AI assistance suggests fixes, compared to traditional code
suggestion technology. Taking inspiration from the
way programmers manually fix bugs, we show that a prompt-based strategy
that conceptualizes repair as localization, transformation, and candidate ranking,
can successfully repair programs in multiple languages with minimal effort.
We present the first results for such a multilingual repair engine
by evaluating on 6 different languages and comparing performance
to language-specific repair engines.
We show that \system{} can outperform language-specific repair
engines for three of these languages.

\end{abstract}

\section{Introduction}
 
The number of people writing code across different languages
has steadily grown~\cite{usbls} and ranges from novices to experts. Regardless of their experience level, programmers can make mistakes when writing code. Program errors can range from those that are easy to spot and fix, to those that require substantial application knowledge and may be very subtle logical bugs. Even simple mistakes, such as syntax errors that require a relatively small edit and may be apparent to a programming expert, can be frustrating for novice programmers.
Moreover they can slow down the workflow of more experienced programmers~\cite{wexelblat1976maxims,murphy2008debugging,altadmri201537,drosos2017happyface}. 
 
One way to help programmers who encounter these small mistakes is by using automated program repair (APR).
These methods take a faulty program and a specification of correctness as input, and return as output fixed version of the program that conforms to the specification. 
Recent work~\cite{lamirage} has introduced the term \emph{last-mile repairs} to broadly describe the class of repairs where the original program is a small edit distance away from the correct program. In this definition, program correctness can be checked without substantial additional context---a parser and a type checker suffice.
A quick search on most programming help forums reveals a large number of questions for such errors.
For example, as of August 2022, there are over 15K posts on StackOverflow tagged with Python and SyntaxError.

Existing work has explored performing these kind of repairs automatically. Symbolic systems,
such as Grmtools~\cite{grmtools}, typically build on error-recovery mechanisms in parsers to enumerate local edits that can resolve errors raised during parsing.
Symbolic systems typically restrict the search space to avoid state explosions and they cannot easily encode properties such as the likelihood of particular repair candidates being correct or not.

More recently, neural approaches have been successfully applied to repairing
syntax and diagnostics errors. For example, Dr. Repair~\cite{drrepair}, BIFI~\cite{bifi}, and TFix~\cite{tfix} use transformer architectures to produce repairs for C compilation errors, Python syntax errors, and JavaScript linter diagnostics, respectively.
Some systems, such as LaMirage~\cite{lamirage}, have also combined symbolic and neural components to successfully repair broken programs in low-code languages such as Excel and Power Fx.

Unfortunately, all these systems share a key drawback: they require substantial engineering (symbolic) or additional data and training (neural) to adapt to new languages.
In this paper, we propose a single repair engine, that leverages a large language model trained on code (LLMC) to perform multilingual repair. We select Codex by OpenAI as the LLMC.

Our system, \system{}, shows that \underline{\textbf{r}}epair \underline{\textbf{i}}s \underline{\textbf{n}}early \underline{\textbf{g}}eneration and exploits Codex's few-shot learning capabilities~\cite{bareiss2022code, drori2022neural} to perform multilingual program repair. To do this effectively, we break down program repair into the same three phases as symbolic automated program repair systems: fault localization, code transformation, and candidate ranking \cite{goues2019automated, liu2021critical, lamirage}. We show how each stage can be addressed with minimal effort by emulating what a developer would do and using this intuition to design prompts for an LLMC.

We evaluate \system{} on six languages: Excel, Power Fx, Python, JavaScript, C and PowerShell.
Our results show that \system{} repairs significantly more programs than a language-specific repair engine for three languages and shows competitive results for another two languages.
We evaluate the effectiveness of our design choices for each of the three stages of repair.
Additionally, we identify possible directions for improvement based on our results, such as language-specific ranking and iterative querying with Codex.

Jointly, these results provide the first evidence that an LLMC can enable multilingual repair with the same or better performance than methods designed for a single language.
In contrast to other AI-assisted code editing features, such as code completion, this advance opens up the possibility of a \emph{flipped interaction model} where the user writes code and the AI assistant performs the fixing.

In summary, we make the following contributions:

\begin{itemize}
    \item We present an LLMC-based approach to multilingual repair that enables a flipped interaction model for AI-assisted programming in which the user writes code and the assistant suggests fixes for last-mile mistakes.
    \item We implement our approach in the \system{} system, which 
    employs compiler (or diagnostic) messages, smart few-shot selection, and ranking of repair candidates to perform repair across varying languages.
    \item We perform an extensive evaluation across six different languages, showing that multilingual repair with LLMCs is viable and can compete with or outperform language-specific repair engines.
    \item We introduce PowerShell commands as a new application for last-mile repair and collect a benchmark set of 200 PowerShell commands from StackOverflow, which
    we also release for future research\footnote{\url{https://github.com/microsoft/prose-benchmarks/}}.
\end{itemize}

\begin{figure*}[ht!]
    \centering
    \includegraphics[width=0.9\textwidth]{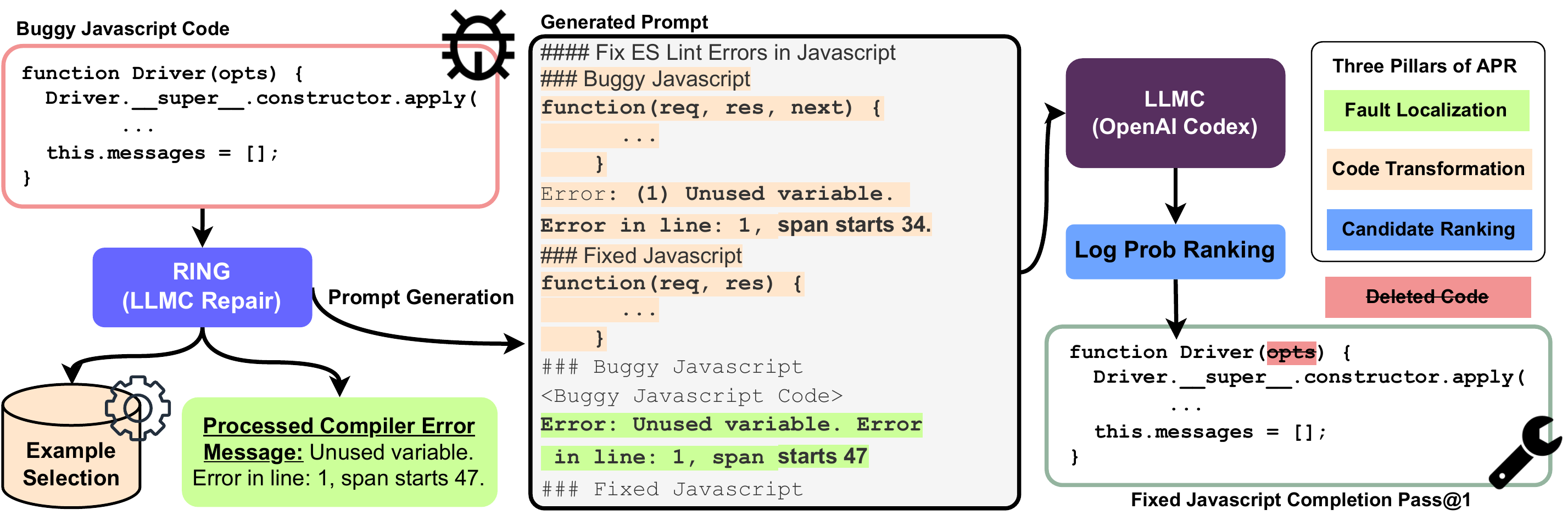}
    \caption{\system{}, powered by a Large Language Model trained on Code (LLMC), performs multi-lingual program repair. \system{} obtains fault localization information from error messages and leverages LLMC's few shot capabilities for code transformation through example selection, forming the prompt. Finally,  a simple, yet effective, technique is used for ranking repair candidates.}
    \label{fig:ring_arch}
\end{figure*}

\section{Related Work}

\paragraph{Automated Program Repair}
Finding and fixing bugs is challenging and tedious, even for language experts \cite{Zhong2015AnES}. The software engineering community has built Automated Program Repair (APR) tools~\cite{arcuri2008automation} to reduce the time and costs associated with debugging.
The premise of APR has since grown into a substantial research domain across different languages, classes of bugs, and use cases \cite{Gazzola2019AutomaticSR}.

Early approaches for APR were symbolic and attempted to fix programs automatically by enumerating repair candidates from templates \cite{Debroy2010UsingMT}, crafting heuristics~\cite{qi2014strength}, and using program synthesis \cite{semfix}. 
Although these systems can provide strong guarantees for the generated code, they are strongly tied to their domain language. Moreover, symbolic systems are restrictive in their scope, failing to repair programs 
that the corresponding language compiler cannot
process to at least some extent. On the other hand, building on the recent advances in natural language processing, neural methods have shown promise in learning program repairs. Researchers have studied automatically correcting programs in different settings,
including introductory programming assignments~\cite{pu2016sk_p, parihar2017automatic, ahmed2018compilation}.
For example, DeepFix~\cite{deepfix} and SampleFix~\cite{samplefix} use sequential deep learning models to fix broken C code written by students. However, these neural models are not as powerful as LLMCs like Codex.

SynFix~\cite{synfix}, Dr. Repair~\cite{drrepair}, and TFix~\cite{tfix} leverage compiler diagnostics for Java, C, and JavaScript, respectively, but require a substantial amount of training and data, failing to generalize across languages. Additionally, neural models can generate plausible but incorrect code, making spurious edits resulting in unparseable code~\cite{guo2021learning}. Although ~\cite{lamirage} tries to bridge the gap between neural and symbolic approaches, it requires language specialization (symbolic parser) and large-scale data (neural localizer and ranker). In contrast, \system{} uses a powerful LLMC, Codex, capable of generating multilingual code while guiding repair through
readily available prompt-design strategies.

\paragraph{Large Language Models}
The advent of Large Language Models (LLM) trained on code and natural language shows promise for code understanding and generation results.
Autoregressive models~\cite{shannon1948mathematical, radford2018improving}, such as Codex~\cite{Codex}, are trained to predict the next token, given the past token context over enormous corpora. However, training LLMs is technically challenging and expensive. They require a large dataset for each fine-tuning task and parallel training on multiple GPUs~\cite{bommasani2021opportunities} to scale.
An exciting aspect of LLMs is the zero-shot and few-shot learning paradigm for adapting to tasks on-the-fly~\cite{gpt3, Chowdhury_Zhuang_Wang_2022}.~\citet{prenner2021automatic} evaluate Codex's ability to fix  80 bugs in Python and Java. They manually provide buggy lines for each program and use \emph{fixed} few shot examples in the prompt. In contrast, our paper discusses various strategies to build the prompt, and performs a more extensive study with larger datasets over more languages.

\section{Approach}

Figure \ref{fig:ring_arch} shows the architecture of \system{}. We divide the task of fixing bugs into three stages: fault localization, program transformation and candidate ranking.
Each stage is based on the intuition for how developers might approach such a stage manually. In the following subsections, we show how to address each stage using an LLMC.

We illustrate our approach using a running example -- shown in Figure~\ref{fig:example-start} -- drawn from the BIFI~\cite{bifi} dataset. The user has incorrectly used tuple notation in the function signature (highlighted in pink). This syntax for unpacking tuples in function signatures was supported in Python 2. In Python 3, it raises a syntax error\footnote{\url{https://peps.python.org/pep-3113/}} with very little detail on the underlying issue.
This example highlights that errors can also be introduced as languages evolve.
\system{} fixes this mistake without additional user intervention.

\begin{figure}
\begin{lstlisting}[language=Python, escapechar=!, basicstyle=\small\ttfamily]
def boundary_difference_power(graph, !\lcolorbox{babypink}{(orig\_image, sigma, spacing)}!):
  orig_image = scipy.asarray(orig_image)
  def boundary_term_division(i):
    i = 1. /(i + 1)
    i = scipy.power(i, sigma)
    i[i <= 0] = sys.float_info.min
    return i
  __skeleton_difference(graph,
        orig_image, 
        boundary_term_division)
\end{lstlisting}
\caption{
A real Python 3 syntax error from the BIFI dataset. 
The highlighted code uses tuple parameter unpacking syntax, which was valid in Python 2 but removed from Python 3. All listings are simplified for presentation clarity and brevity.}
\label{fig:example-start}
\end{figure}

\subsection{Fault Localization through Language Tooling}

As a first step towards debugging, a programmer typically locates the cause of the bug.
For most modern languages, locating syntactic mistakes and some semantic errors, such as type errors, is aided by tools like the compiler, static analyzers, or linters.
Following this intuition, we include a preprocessed error message produced by the compiler or other static analyzers. We normalize this message to enforce consistency across languages.
Figure~\ref{fig:compiler-message-prompt} shows this prompt variant for our running example, where the highlighting corresponds to our prepared syntax error message. 
For languages where the error messaging may not be precise, particularly with regards to the error location reported, we found that a simple abstraction that removes the reported error location but preserves the error text worked well -- we discuss how to create such an abstracted message in our discussion section.

\begin{figure}
\begin{lstlisting}[language=Python, escapechar=!,basicstyle=\small\ttfamily]
### Buggy Python
def boundary_difference_power(graph,
    (orig_image, sigma, spacing)):
...
!\lcolorbox{babypink}{\small Error: (1) invalid syntax. Error in}!
!\lcolorbox{babypink}{line: 2 span starts 4 and ends 32.}!
\end{lstlisting}
\caption{
To aid fault localization, we include a detailed compiler error message with line/column span information. We prepare uniform messages across languages by extracting details from the corresponding language compiler/analyzer.
}
\label{fig:compiler-message-prompt}
\end{figure}

\subsection{Code Transformation through Few-shot Learning}

Once a developer has identified the location of a mistake, they must now apply an appropriate transformation---a sequence of edits---to the original source code at this location.
Most developers accumulate experience in the type of transformations needed to resolve particular errors over time.
Additionally, when novices encounter an unfamiliar mistake, they often search for examples of similar buggy/correct pairs that can inform their own transformation.

It has been shown that LLMs are capable of few-shot learning---the ability to learn from a few examples of the intended task---by adding related examples of the task to the prompt \cite{gpt3, synchromesh}.
Given examples of transformations that repair programs, we exploit this capability in \system{} to address the code transformation stage. The main challenge is selecting relevant examples that are related to the mistake made by the developer.

Following the intuition that programs with similar mistakes have similar fixes, we select examples from a collection of buggy-fixed pairs based on error message similarity.
We call this collection of buggy-fixed pairs the example bank.

To capture differences in language tooling, we implement two methods for selecting programs from our example bank. The key difference between these two methods is how they compute a similarity metric over
error diagnostics.

The first variant, \emph{error vector selection}, assumes that fine-grained error reporting is available.
For example, the Excel parser returns a detailed report with many different diagnostic counters.
We count the occurrence of each error
category reported by the tool and construct
a vector out of these frequencies -- we refer to this as an error vector. We then select programs from the example bank by minimizing the L2 distance between error vectors.

The second variant, \emph{message embedding selection}, assumes that high-level errors are accompanied by detailed descriptions in natural language.
For example, the Python parser often returns the same error (like SyntaxError) for different mistakes and instead exposes additional information through the associated natural language error message.
We use this description by embedding the compiler messages with a pre-trained CodeBert~\cite{feng2020codebert} model and comparing embeddings based on cosine similarity.

Figure~\ref{fig:smart-shots-prompt} shows a simplified few-shot prompt with an example, chosen using message embedding, which exhibits the same error (and required fix) as our buggy program.
With this prompt, \system{}'s top candidate is the right repair.

\begin{figure}[htb]
\begin{lstlisting}[language=Python, escapechar=!, basicstyle=\small\ttfamily]
!\footnotesize\lcolorbox{babypink}{\#\#\# Buggy Python}!
!\footnotesize\lcolorbox{babypink}{\textbf{def} initial\_solution(self, start,}!
    !\footnotesize\lcolorbox{cinnabar}{\textbf{(max\_shares, desired\_weight)}}!!\colorbox{babypink}{):}!
!\footnotesize\lcolorbox{babypink}{...}!
Error: (1) invalid syntax. Error in line: 3, span starts 35 and ends: 36.
!\footnotesize\lcolorbox{babypink}{\#\#\#  Fixed Python}!
!\footnotesize\lcolorbox{babypink}{def initial\_solution(self, start,}!
    !\footnotesize\colorbox{celadon}{\textbf{max\_shares, desired\_weight}}!!\colorbox{babypink}{):}!
!\footnotesize\lcolorbox{babypink}{...}!
\end{lstlisting}
\caption{
Our \emph{smart selection of few-shots} 
retrieves relevant buggy-fix examples from an example bank.
Shots are retrieved based on a similarity metric over error diagnostics. The shot selected (pink background) displays the same invalid signature-level tuple parameter unpacking (dark red background, \textbf{bold}) as our target program.
The fixed portion of the shot (green background, \textbf{bold}) removes the parentheses.
}
\label{fig:smart-shots-prompt}
\end{figure}

\subsection{Candidate Ranking}

LLMs achieve variation in their output by iteratively sampling each token from promising candidates.
The extent to which less likely tokens can be selected is controlled by a parameter called \emph{temperature}. 
We can thus generate multiple candidates by controlling the temperature during generation. 

The final step in \system{} is to rank the candidates obtained by querying Codex using the prompt described in the prior two stages. We use a relatively simple (but effective) ranking strategy to order the candidate programs: averaging the log-probabilities of tokens selected during the decoding process and sort the candidates in descending order of their averages.

During development, we found that generating various candidates with higher temperatures -- encouraging diverse candidates -- and ranking them yields better performance than using lower temperatures such as zero. 

\section{Language-Specific Datasets}

We evaluate \system{} on six different languages, ranging from low-code formula languages to popular scripting languages. We describe the dataset, language-specific baseline(s) and evaluation metric for each language.

\paragraph{Excel} 

We use a recently released dataset of 200 Excel repair tasks collected from Excel help forums \cite{lamirage}.
Each task consists of an Excel formula with syntax errors, some semantic errors (such as wrong function call arity) and a ground truth repair.
We also collect a set of 73 tasks where the Excel formula contains at least one type error and annotated each such formula with a ground truth repair. The final collection consists of 273 Excel repair tasks.

A successful repair exactly matches the ground truth after normalizing tokens like spaces, capitalizing all the identifiers and cell references.
We compare \system{} to the neurosymbolic repair engine LaMirage \cite{lamirage}.

\paragraph{Power Fx}

Like Excel, we use the recently released 200 Power Fx repair tasks accompanying LaMirage. These tasks consist of syntactic and basic semantic errors, and are collected from help forums and anonymized product telemetry.

We use the same evaluation criteria as in Excel and compare to the neurosymbolic repair engine LaMirage.

\paragraph{Python}

We evaluate \system{} on a random sample of 200 syntactically invalid Python code snippets from the dataset used by the SOTA syntax repair tool for Python: BIFI~\cite{bifi}.
These code snippets were collected from GitHub repositories.

These snippets do not have a ground truth repair, hence, we employ the same evaluation metric described in the BIFI paper.
A repair is successful if the produced program is (1) parsed successfully by the Python 3 parser and (2) has a Levenshtein~\cite{levenshtein1966binary} token edit distance less than 5 from the buggy program.
The pyhon tokens are generated by the Pygments\footnote{\url{https://pygments.org/}} lexer.

We compare to BIFI, a transformer-based repair system that iteratively trains a \emph{code breaker} that learns to generate realistic errors and a \emph{code fixer} that repairs such errors.

\paragraph{JavaScript}

We evaluate \system{} on a random sample of 200 JavaScript (JS) code snippets drawn from the dataset released with TFix~\cite{tfix}.
Each snippet has at least one error or warning reported by the popular linter ESLint~\cite{eslint}.
In addition to syntax errors, ESLint also reports stylistic issues.

The dataset released by TFix contains a ground truth repair code snippet for each buggy snippet.
Both buggy and ground truth code snippets were mined by the TFix authors from GitHub commits.
The originally released dataset contains only the part of each code snippet relevant to the error and repair.
However, these parts are an arbitrary window around the original fault location.
We found that providing these arbitrary windows to Codex resulted in false edits, as the windows had syntax errors that were just an artifact of the windowing.
To mitigate this, we extracted the whole function (or whole file, if not in a function) that encompassed the originally buggy and the repaired code snippets.
We refer to these as \emph{extended code snippets}.

We compare our performance to TFix, a fine-tuned T5~\cite{t5} model for JS repair.
A repair is successful if it matches the ground truth corresponding to the buggy program.
We run TFix on both the original window snippets and on our extended code snippets.

\paragraph{C}

We evaluate \system{} on a random sample of 200 C code snippets drawn from the dataset released with DeepFix \cite{deepfix}.
These programs correspond to real user programs written by students in an introductory programming class and raise at least one compilation error.

We compare to Dr. Repair, a neural repair system that uses graph attention to combine information from the buggy code snippet and the associated compiler message \cite{drrepair}.
We use their success criterion: a repair must not raise any error messages when compiled using \texttt{gcc -w -std=c99 -pedantic}.
Following BIFI, a repair must be less than 5 token edits away from the original buggy program.

\paragraph{PowerShell}

We introduce the novel task of repairing syntax errors in PowerShell commands.
To create benchmarks, we searched StackOverflow \cite{stackoverflow} for the word ``error'' in threads tagged with \texttt{powershell}.
This resulted in 14,954 threads. We extracted code blocks with least one space from the question and the accepted answer.
We keep pairs from question and answer where the question code is invalid and answer code is valid.
We judged validity using the PowerShell command \lstinline{Get-Command -syntax}. 

Finally, we manually annotated these candidate tasks from the the associated StackOverflow post, confirming each pair was reflective of the original issue and did not have extra changes.
We kept a final set of 208 task pairs.

There is no existing language-specific engine to compare with, as we introduce this task.
A repair is successful if it exactly matches the associated answer code block.

\paragraph{Common Baseline}

We also use zero-shot Codex as a baseline for all languages.
We use the following prompt: 
\begin{verbatim}
Fix bugs in the below code:
### Buggy <language>: 
<buggy program>
### Fixed <language>:
\end{verbatim}
where \texttt{<language>} is replaced with the appropriate language name for the benchmark task.

\begin{table*}[t!]
\small
\centering
\begin{tabular}{llrrrcr}
\toprule
\textbf{Language} & \textbf{Approach} & \multicolumn{1}{r}{\textbf{Pass@1}} & \multicolumn{1}{r}{\textbf{Pass@3}} & \textbf{Pass@50} & \multicolumn{1}{c}{\textbf{Metric}} & \multicolumn{1}{r}{\textbf{Avg. Tokens}} \\ \midrule
\multirow{3}{*}{Excel} & \system{} (Abstracted Message, Error Vector) & \textbf{0.82} & \textbf{0.89} & \multicolumn{1}{r}{\textbf{0.92}} & \multirow{3}{*}{Exact Match} & \multirow{3}{*}{26 $\pm$14} \\ 
&LaMirage~\cite{lamirage} & 0.71 & 0.76 & - & & \\ 
&Codex~\cite{Codex}       & 0.60 & 0.77 & 0.88 & & \\ \midrule
\multirow{3}{*}{Power Fx} & \system{} (Compiler Message, Message Embedding) & 0.71 & 0.85 & \multicolumn{1}{r}{\textbf{0.87}} & \multirow{3}{*}{Exact Match} & \multirow{3}{*}{29 $\pm$19} \\ 
                         & LaMirage~\cite{lamirage} & \textbf{0.85} & \textbf{0.88} & - & & \\
                         & Codex~\cite{Codex}& 0.47 & 0.68 & 0.84 & & \\ \midrule
\multirow{4}{*}{Javascript} & \system{} (Compiler Message, Error Vector) & 0.46 & 0.59 & \multicolumn{1}{r}{\textbf{0.64}} & \multirow{4}{*}{Exact Match} & \multirow{4}{*}{163 $\pm$106} \\ 
                            &TFix (extended code snippets) \cite{tfix} & 0.09 & \multicolumn{1}{r}{-} & - & & \\
                            &TFix (original dataset) \cite{tfix} & \textbf{0.59} & \multicolumn{1}{r}{-} & - & & \\
                            &Codex~\cite{Codex}& 0.19 & 0.28 & 0.39 & & \\ \midrule
\multirow{3}{*}{Python} & \system{} (Compiler Message, Message Embedding) & \textbf{0.94} & \textbf{0.97} & \multicolumn{1}{r}{0.97} & \multirow{3}{*}{\shortstack[c]{Passes Parser\\ Edit Distance $<$ 5}} & \multirow{3}{*}{104 $\pm$150} \\
                       & BIFI~\cite{bifi}   & 0.92 & 0.95 & 0.96 & & \\
                       & Codex~\cite{Codex} & 0.87 & 0.94 & \textbf{0.98} & & \\ \midrule
\multirow{3}{*}{C} & \system{} (Compiler Message, Message Embedding) & \textbf{0.63} & \textbf{0.69} & \multicolumn{1}{r}{\textbf{0.70}} & \multirow{3}{*}{\shortstack[c]{Passes Parser\\ Edit Distance $<$ 5}} & \multirow{3}{*}{223 $\pm$72} \\
                   & Dr Repair~\cite{drrepair} & 0.55 & - & - & & \\ 
                   & Codex~\cite{Codex}& 0.40 & 0.56 & 0.61 & & \\ \midrule
\multicolumn{1}{c}{\multirow{2}{*}{Powershell}} & \system{} (Compiler Message, Message Embedding) & \textbf{0.10} & \textbf{0.19} & \multicolumn{1}{r}{\textbf{0.27}} & \multirow{2}{*}{Exact Match} & \multirow{2}{*}{24 $\pm$30} \\
                   % & No language-specific baseline & - & - & - & & \\
                   &Codex~\cite{Codex}& 0.06 & 0.11 & 0.18 & & \\ \bottomrule
\end{tabular}
\caption{Comparison of \system{} with language-specific approaches and a zero-shot baseline that uses Codex. \textbf{Bold} denotes best performance for each language. All \system{} experiments are at 0.7 temperature.
	\system{} can outperform language-specific repair engines in Excel, Python, and C.
	In Javascript, \system{} is capable of generating the right repair but
	ranking needs to improve. In Powershell, with no existing baseline, \system{} performs substantially worse -- likely reflective of the lack of Powershell code in Codex's training data. 
	We ran all Codex-related queries on 9\textsuperscript{th} August 2022 using Open AI's public API for ``davinci-code-002".
}
\label{table:main_results}
\end{table*}

\section{Results and Analysis}

We first ask: (\textbf{RQ1}) how viable is \system{}'s Codex-powered approach for repair across multiple languages?
Next, we investigate the extent to which Codex can address each of our conceptual stages.
For localization, (\textbf{RQ2}) to what extent can \system{} perform error localization across languages?
For code transformation, (\textbf{RQ3}) to what extent does our smart selection of few-shots improve performance?
Finally, for candidate ranking, (\textbf{RQ4}) to what extent can \system{} rely on Codex's token log probabilities to rank candidates?

\subsection{RQ1. Viability of Multilingual Repair}

Table~\ref{table:main_results} shows  the performance for \system{}, language-specific repair engines, and a Codex-based zero-shot baseline, across each of our languages.
We present the best performing configuration for each language using \emph{pass@k} performance metrics~\cite{inala2022fault,synchromesh,lamirage}.

Smart selection is done via leave-one-out. For languages with ground truth, all other tasks are the example bank for drawing shots.
Since the C and Python datasets do not have ground truth pair, we sample additional 400 programs from their corresponding datasets.
We run the best RING configuration (without smart selection) on these 400 programs and pick those that do not raise any diagnostics error.
These buggy/correct pairs form the example bank in C and Python.

\system{} outperforms the state-of-the-art repair engines in pass@1 for Excel, Python, and C.
For Power Fx, we find that \system{}'s pass@3 rate is comparable to the pass@1 rate for LaMirage.
Furthermore, there is a substantial improvement in \system{}'s pass@3 compared to pass@1.

In Javascript, we find that TFix applied to the original code snippets obtains a pass@1 rate of 0.59 (approximately
7 points higher than that of \system{}).
However, applying TFix to the extended code snippets results in a much lower pass@1 rate of 0.09.
This performance degradation can be attributed to the substantially longer sequences of the extended code snippets compared to the original code snippets, an average of 208 and 74 T5 tokens, respectively.

In PowerShell (PS), we observe that \system{}'s performance is substantially lower compared to other languages.
We hypothesize that this may be a reflection of the (presumed) relative scarcity of PS commands in Codex's training data.
Manual inspection of failures also revealed that \system{} performs fewer edits than required to match the ground truth.

Given this evidence, we conclude that \system{}'s Codex-powered approach can perform multilingual repair.
We can contrast this to the substantial effort required to build a language-specific repair engine.
TFix, BIFI, and Dr. Repair were trained on 108K, 3M and 1.5M JavaScript, Python, and C code snippets, respectively.
LaMirage trained error localizers and rankers based on pointer networks, as well as implemented multiple language-specific rules.

Programs that were not fixed by either \system{} or the language-specific engines shared some properties.
In particular, some of these could be addressed with a combination of iteratively querying Codex and explicit lightweight constraints that enforce language-specific knowledge.
For example, we found a Python program that has two issues: an invalid use of the reserved keyword \lstinline|async| and a missing parenthesis.
For the keyword issue, we could query Codex with the buggy program up to the invalid keyword usage, validate that the following token predicted is not a reserved keyword, and then query Codex \emph{again} with the modified buggy code fragment.
This is similar to constrained decoding used in Synchromesh~\cite{synchromesh}.

\subsection{RQ2. Error Localization}

Even if \system{} cannot fix a program at pass@1, locating the error can help users.
We carry out the following experiment for the four languages which have the ground truth.
We consider programs that are not repaired at pass@1 by \system{} and those that are not repaired at pass@1 by the language-specific baseline.
For each such program, we take the top candidate produced by each system and compare the edit locations to the ground truth edit locations.
If the candidate edit locations are all within a range of $\pm$k tokens of the ground truth locations, we mark this as a correct localization.

\begin{figure}[h!]
    \centering
    \small
    \includegraphics[width=1\linewidth, height=4.5cm]{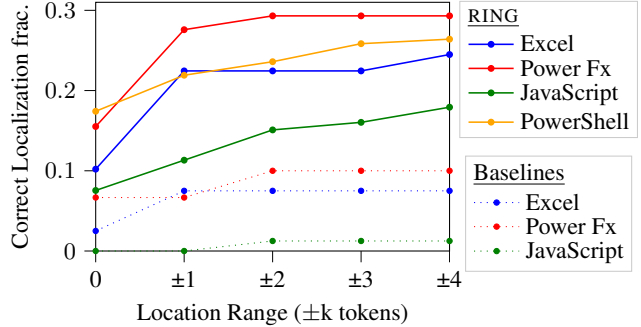}
    \caption{
    We consider separately the programs not repaired
    at pass@1 by \system{} and 
    language-specific baselines. We compute an approximate error localization metric, which marks as correctly localized any edit that is within $k$ tokens of the groundtruth edit location. When \system{} fails to repair a program it correctly localizes a larger fraction of programs compared to the language-specific baselines.
    }
    \label{fig:edit-location}
\end{figure}

Figure~\ref{fig:edit-location} summarizes our results.
We observe that \system{} correctly locates a larger fraction of required edits compared to the language-specific baselines.
This holds true across the four languages with ground truth repairs.
\system{}'s localization success varies by language but can reach as high as over a quarter of unrepaired programs (for Power Fx, given a tolerance of one token).
For such programs, where \system{} can localize the error but does not perform the correct edit, drawing shots from a larger example bank may help.

\begin{figure}
\centering
\begin{subfigure}{\linewidth}
\small
    \includegraphics[width=1\linewidth, height=3.5cm]{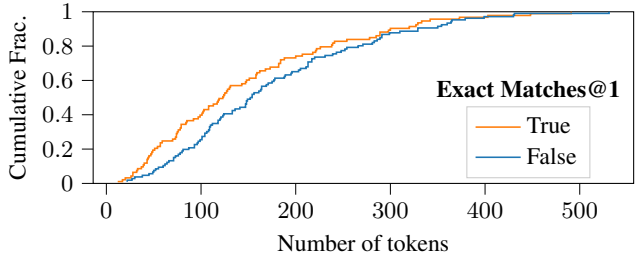}
    \caption{Javascript}
\end{subfigure}
\\
\begin{subfigure}{\linewidth}
\small
    \includegraphics[width=1\linewidth, height=3.5cm]{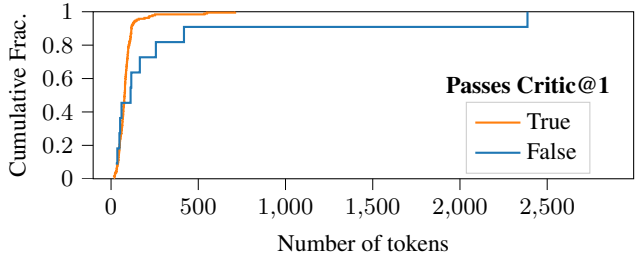}
    \caption{Python}
\end{subfigure}
\caption{
Cumulative fraction of programs by number of tokens in the original buggy program, grouped by whether \system{} can repair at pass@1. Successful repairs tend to be
associated with shorter buggy programs.}
\label{fig:program-length}
\end{figure}

Next, we explored a key contributing factor to overall repair success (and localization in particular): program length.
We found that for most languages, the buggy programs that \system{} can repair tend to be shorter than the buggy programs it fails to repair
Figure~\ref{fig:program-length} shows the cumulative fraction of buggy programs by their length, grouped based on their outcome (pass@1).
In both JavaScript and Python, the programs successfully repaired by  \system{} tend to be shorter than those where it fails.
Interestingly, this relationship does not seem to hold as strongly for Excel.
We attribute this behaviour to shorter programs lengths and the restrictive Excel grammar.

\begin{figure}[t!]
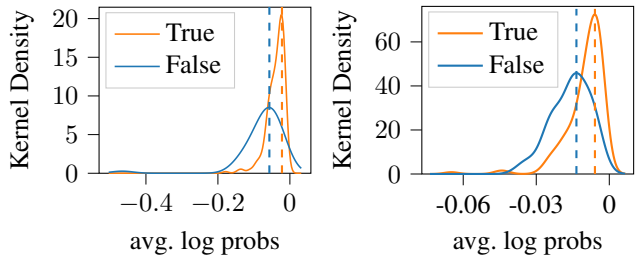
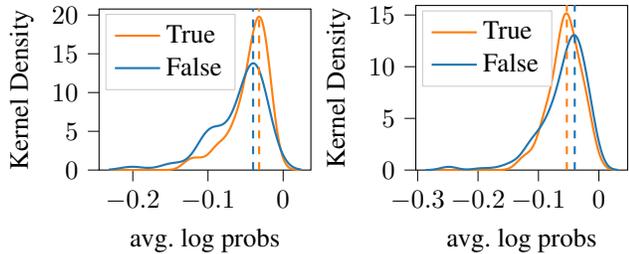

\begin{subfigure}{0.49\columnwidth}
    \includegraphics[width=1\linewidth, height=3.5cm]{figures/excel_logprobs.tikz}
    \caption{Excel}
\end{subfigure}
\begin{subfigure}{0.49\columnwidth}
    \includegraphics[width=1\linewidth, height=3.5cm]{figures/c_logprobs.tikz}
    \caption{C}
\end{subfigure}
\begin{subfigure}{0.49\columnwidth}
    \includegraphics[width=1\linewidth, height=3.5cm]{figures/powerapps_logprobs.tikz}
    \caption{Power Fx}
\end{subfigure}
\begin{subfigure}{0.49\columnwidth}
    \includegraphics[width=1\linewidth, height=3.5cm]{figures/powershell_logprobs.tikz}
    \caption{Powershell}
\end{subfigure}
\caption{
(Gaussian) Kernel density plots for average token log probabilities
across languages, based on their pass@1 success status. Clearer separation of distributions
tends to be associated with better performance (e.g., Excel, C). In Powershell, where \system{} struggles, the relationship between distribution peaks is inverted relative to other languages.}
\label{fig:log-probs-by-outcome}
\end{figure}

\subsection{RQ3. Code Transformation}

Table~\ref{tab:few-shots} shows the pass@1 rate with our smart selection of few-shots for the prompt, compared to a strategy that uses pre-defined fixed examples.
The pre-defined strategy allows us to curate high-quality repair examples for common errors, but these may not be relevant for all programs.
Prior work~\cite{prenner2021automatic} explored the use of fixed few-shot examples for APR with Codex.

Our results show that smart selection improves performance in all languages.
This performance improvement comes from examples in the prompt that reflect similar errors (and expected edits) to the target program.
We use error vector selection for Excel and JavaScript, which have better and more granular error categorization, and message embedding selection for other languages.
We observe that Power Fx shows the smallest performance improvement.
Manual inspection revealed that Power Fx compiler messages tend to be imprecise, and using them to select examples can introduce some noise into the prompt.
An example that we encountered were cases where the compiler suggested there was an extraneous token in the input program that did not actually appear in it.

\begin{table}[t!]
\centering
\small
\begin{tabular}{@{}lrrr@{}}
\toprule
Language     & Fixed Shots & Smart Shots & Fractional Change \\ \midrule
Excel      & 0.76        & 0.82        &  0.08 \\
Power Fx    & 0.70        & 0.71        &  0.01   \\
Javascript & 0.43        & 0.46        &  0.07 \\
Python     & 0.91        & 0.94        &  0.03 \\
C          & 0.5         & 0.58        &  0.16 \\
Powershell & 0.09        & 0.1         &  0.01 \\ \bottomrule
\end{tabular}
\caption{
Pass@1 for few-shots selected using our smart selection
strategy, compared to pre-defined fixed examples. Smart selection improves performance for all languages. For Power Fx, we see the smallest improvement, which we attribute to imprecise compiler diagnostics.
}
\label{tab:few-shots}
\end{table}

\subsection{RQ4. Candidate Ranking}

\system{} ranks candidate repairs based on the average of per-token log probabilities produced by Codex.
The effectiveness of this strategy for our use case depends on the extent to which Codex is calibrated properly for program repair \cite{bella2010calibration,nixon2019measuring, dormann2020calibration}.
Figure~\ref{fig:log-probs-by-outcome} compares average log probabilities in Excel, Power Fx, PowerShell, and C.
We show (Gaussian) kernel density plots across languages based on the pass@1 outcome.
For languages like Excel and C, where \system{} outperforms language-specific repair engines, there is a clearer difference in distributions.
In Power Fx, where \system{} can repair programs but does not outperform the language-specific engine, this distribution
difference is less clear.
In PowerShell, where \system{} fails to repair a substantial fraction of programs, the relationship between the peaks of the distributions is inverted relative to other languages.

\begin{figure}[ht!]
    \centering
    \includegraphics[height=3.5cm, width=\linewidth]{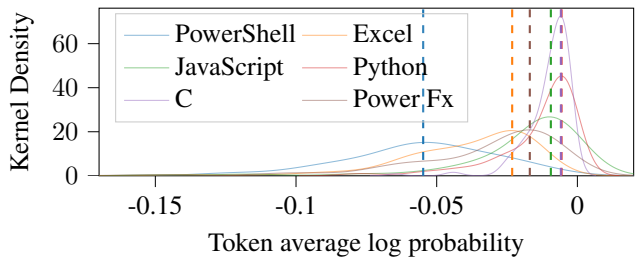}
    \caption{
    Per-language
    (Gaussian) kernel density plots of
    successful pass@1 scores (average token log probabilities).
    We find that less popular languages, like PowerShell, Excel, and Power Fx,
    have lower average scores -- likely reflective of their relatively small fraction of Codex's training data.}
    \label{fig:pass-at-1-logprobs}
\end{figure}

Additionally, we find that even for programs with a pass@1 success outcome, there are differences in average token log probability across languages, as shown in Figure~\ref{fig:pass-at-1-logprobs}.
For less popular languages (PowerShell or Excel), the distribution peaks are further left than more popular languages (JavaScript). This likely reflects the underlying language distribution in Codex's training data.

Based on our observation of the gap between pass@1 and pass@5, paired with these calibration insights, we believe that a dedicated ranking model~\cite{inala2022fault} may provide a substantial payoff in multilingual repair.

\section{Discussion}
We now provide discussion on the design principles involved in building a good
example bank for few-shot selection
and the tasks required to adapt
\system{} to a new language.

\subsection{Designing the Example Bank}

While curating the example bank, it is essential to have different types of errors to facilitate retrieval of similar mistakes/fixes for a new buggy program.
There are several ways to collect such examples, including scraping public forums, using telemetry data, and bootstrapping examples through the language knowledge of an expert.
We have found that scraping public forums is a good way to start, paired with expert curation of corner cases.
Buggy-fixed pairs can be collected incrementally, adding more diverse examples from different sources later.
Telemetry data also provides a natural source for examples, but depending on the platform/organization can require anonymization that might impact retrieval.

Our evaluation employs a strict leave-one-out strategy to build an example bank from  benchmark programs.
In practice, this will be a very restrictive example bank that can potentially limit the number of successful repairs. 

While the example bank sizes used during our evaluation do not present a performance concern, as example banks in production grow, retrieval time may become more significant.
To address such challenges, \system{} could take advantage of off-the-shelf fast indexing/retrieval systems, such as
FAISS~\cite{faiss} or ANNOY~\cite{annoy}.

\subsection{Adapting \system{} for new languages}

We now detail the steps required to apply \system{} to a new language.
The first task is to build the associated example bank, using the principles discussed above.
Next, we need to evaluate the language tooling available for error diagnostics.
In particular, there are two key decisions: determining what kind of error-based few-shot selection to make and if the error message needs to be abstracted prior to use for localization with \system{}. 
We discuss each of these concerns in turn.

\paragraph{Choosing between Error Vector and Message Embedding}
Different languages have different underlying language tools, such as compilers and linters.
If the underlying language tools provide detailed error reports with granular error categories, counting the categories can help us extract precise error information.
We recommend using error vector selection for such languages.
Unfortunately, not all languages provide fine-grained error categories but instead expose additional information through an associated natural language error message. 
For such languages, which provide more information through natural language, we recommend using message embedding selection.

\paragraph{Creating abstracted error message}

When incorporating the error message in the localization portion of the prompt and in message-embedding-based few shot selection, some languages may benefit from  abstracting the error message to remove extra (and possibly imprecise) information.
In our experiments, we found that providing an error message without exact location information can help in low-code languages like Excel.
If the language tool provides data structures with error description, location, and error category, we only use the description. 
Languages with natural language error messages typically follow a template that we can use to extract the portions of the message we want to preserve to create the abstracted message.

For example, in C, the error message for a missing semicolon (;) at the end of a statement is shown below:
\begin{verbatim}
In function 'main':
16:6: error: expected ';' before 'printf'
      printf(\"%d\",catalan(h));
      ^
\end{verbatim}
We split the error message using regular expression ``\texttt{\textbackslash d+:\textbackslash d+: error:}", which captures the text \texttt{16:6 error:} and leaves us with the following abstracted error message: \texttt{expected ';' before 'printf'}.

\section{Conclusion}
We present \system{}, a multilingual repair engine
powered by Codex.
We show various prompt-based strategies designed to 
convey developer-like information to address
the three stages of automate program repair: 
error localization, code transformation, and
candidate ranking. We evaluate \system{}
on six languages, including a benchmark for the
novel task of repairing Powershell programs.
We show \system{}
can perform well in multiple languages, even outperforming
language-specific engines in some, with little engineering effort.

\section{Acknowledgements}

We thank Peter Lee for the CoPilot-flipped-model analogy. We thank Julia Liuson
for inspiring and facilitating use of Codex-as-a-component in our
neuro-symbolic workflows. We also thank the authors of 
baseline systems used in our research -- their sharing of models
and data made this work possible.
We would also like to thank Abishai Ebenezer for helping us curate the PowerShell evaluation benchmarks.

\bibliography{ref}

\clearpage 

\appendix

\section{Appendix}

\subsection{Fewer edits in PowerShell}
\begin{figure}[ht!]
    \centering
    \includegraphics[width=0.7\linewidth]{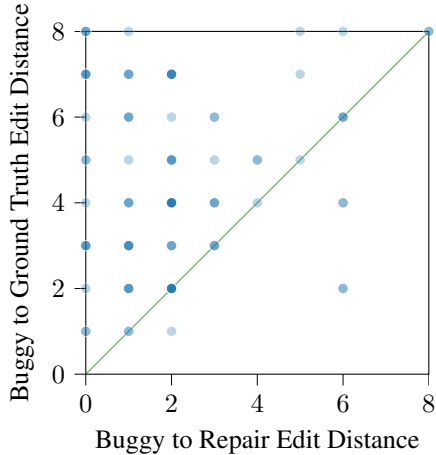}
    \caption{For incorrect Pass@1 repair attempts for Powershell, we observe that Codex makes fewer edits than required, based
	    on the edit-distance differences between
	    predicted and ground-truth programs
	    when compared to the original
	    buggy program.}
    \label{fig:ps_repair_vs_gt}
\end{figure}

Manual inspection of failures also revealed that \system{} performs fewer edits than required to match the ground truth.
as shown in Figure~\ref{fig:ps_repair_vs_gt},
which compares the edit distance between
the buggy and predicted repaired program and the buggy and groundtruth correct program.

\subsection{Combining \system{} with language-specific repair engines}
Another possibility for further improving
the performance of \system{}, is to 
combine the candidate repairs it suggests with those produced by language-specific repair engines -- a method akin to ensembling~\cite{sagi2018ensemble}. We found that at pass@1, the language-specific repair engines for Power Fx and Javascript can repair
34 programs (17\%) and
52 programs (26\%), respectively, that \system{} does not repair.
The possibilities for complementary
repairs are fewer in Excel (15 formulas), Python (6 programs), and C (4 programs). 

\subsection{Program Normalization}
In this section we detail normalization steps taken for candidate dedeuplication and matching to ground truth. We do not perform normalization for Powershell and PowerFx. For generated candidates, we filter out those which exactly match the buggy code. We carry out these normalization steps for all systems, both \system{} and baselines.

\paragraph{Excel}
\begin{itemize}
    \item Excel tokens are case sensitive, therefore, we capitalize all the cell references and identifier.
    \item Remove whitespace tokens
\end{itemize}

\paragraph{Python}
\begin{itemize}
    \item Remove comments
    \item Remove redundant new lines
\end{itemize}

\paragraph{Javascript}
\begin{itemize}
    \item Remove whitespace
    \item Remove new line
    \item Remove comments
\end{itemize}

\paragraph{C}
\begin{itemize}
    \item Remove comments
    \item Remove whitespace
    \item Remove new lines
\end{itemize}

\begin{figure}[t!]
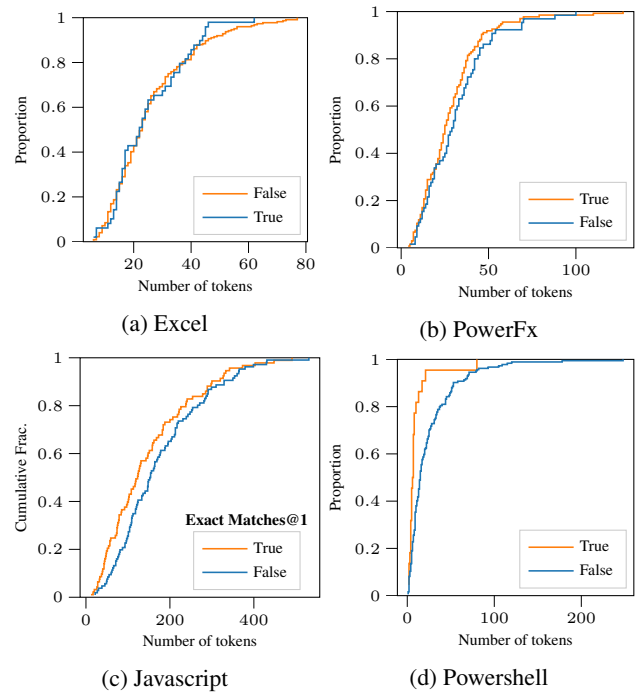

\tiny
\centering
\begin{subfigure}{0.49\columnwidth}
    \includegraphics[width=1\linewidth]{figures/num_tokens_cdf_excel.tikz}
    \caption{Excel}
        \label{fig:excel_num_tokens_cdf_supp}
\end{subfigure}
\begin{subfigure}{0.49\columnwidth}
    \includegraphics[width=1\linewidth]{figures/num_tokens_cdf_powerapps.tikz}
    \caption{PowerFx}
        \label{fig:powerfx_cdf_supp}
\end{subfigure}
\begin{subfigure}{0.49\columnwidth}
    \includegraphics[width=1\linewidth]{figures/num_tokens_cdf_js.tikz}
    \caption{Javascript}
        \label{fig:js_num_tokens_cdf_supp}
\end{subfigure}
\begin{subfigure}{0.49\columnwidth}
    \includegraphics[width=1\linewidth]{figures/num_tokens_cdf_powershell.tikz}
    \caption{Powershell}
        \label{fig:ps_num_tokens_cdf_supp}
\end{subfigure}
\caption{
Cumulative fraction of programs, grouped by pass@1 outcome,
by number of program tokens in the original buggy program.
Languages with groundtruth repair available.}
\label{fig:supp_number_of_tokens_matches_critic}
\end{figure}

\begin{figure}[t!]
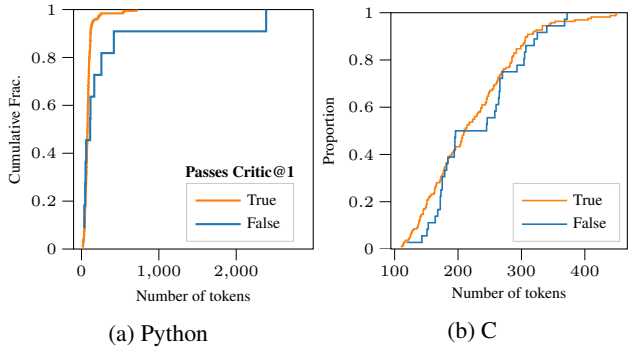

    \centering
    \tiny
    \begin{subfigure}{0.49\columnwidth}
    \includegraphics[width=1\linewidth]{figures/python_num_tokens_cdf.tikz}
    \caption{Python}
        \label{fig:python_num_tokens_cdf_supp}
\end{subfigure}
\begin{subfigure}{0.49\columnwidth}
    \includegraphics[width=1\linewidth]{figures/c_num_tokens_cdf.tikz}
    \caption{C}
        \label{fig:c_num_tokens_cdf_supp}
\end{subfigure}
\caption{Cumulative fraction of programs, grouped by pass@1 outcome,
by number of program tokens in the original buggy program. Languages without groundtruth repair.}
\label{fig:supp_number_of_tokens_critic_pass}
\end{figure}

\subsection{Number of Tokens vs Number of repairs}
In Figure~\ref{fig:supp_number_of_tokens_matches_critic}, we observe that for most languages, programs that \system{}
successfully repairs tend to be shorter
than those it fails to repair.
 Interestingly, for Javascript the gap between proportion of programs decreases as number of tokens increase. In Excel, there is no clear distinction, likely due to smaller program length in the dataset compared to other languages.

In Figure~\ref{fig:supp_number_of_tokens_critic_pass}, we observe similar trend for Python, but in C we find some overlap.

\subsection{Average log probability}
We observe that for all languages except Powershell, \system{} is more confident about correct repairs, as shown in main paper Figure 7 and, Figure,~\ref{fig:python_logprobs},~\ref{fig:js_logprobs},~\ref{fig:c_logprobs}.

\begin{figure}[h!]
\centering
\begin{subfigure}{0.49\linewidth}
    \includegraphics[width=1\linewidth, height=4cm]{figures/python_logprobs.tikz}
    \caption{Python: Average token log probabilities density}
        \label{fig:python_logprobs}
\end{subfigure}
\begin{subfigure}{0.49\linewidth}
        \includegraphics[width=1\linewidth, height=4cm]{figures/js_logprobs.tikz}
    \caption{JS: Average token log probabilities density}
        \label{fig:js_logprobs}
\end{subfigure}
\begin{subfigure}{1\linewidth}
            \includegraphics[width=1\linewidth, height=4cm]{figures/c_logprobs.tikz}
    \caption{C: Average token log probabilities density}
        \label{fig:c_logprobs}
\end{subfigure}
\end{figure}

\subsection{No token edit limit}
In the results section of the main paper, we report \system{} results with a token edit distance threshold of 4. In Table~\ref{table:c_and_python_without_token_edit_dist}, we present results without the token edit distance. We observe that \system{} still outperforms BIFI~\cite{bifi} and Dr Repair~\cite{drrepair} in Python and C, respectively.

\section{Different Configurations of \system{}}
From Table~\ref{table:config_excel} to~\ref{table:config_c}, we present results of \system{} with different configurations. For Excel and PowerFx, we see that Abstracted compiler message gives better performance, likely due to shorter (usually, 1 line) programs.

\begin{table}[t!]
\small
\centering
\begin{tabular}{llrrr}
\hline
\multicolumn{1}{c}{}                                        & \multicolumn{1}{c}{}                                          & \multicolumn{3}{c}{\textbf{Metrics: Exact Match}}                                                                \\ \cline{3-5} 
\multicolumn{1}{c}{\multirow{-2}{*}{\textbf{Localization}}} & \multicolumn{1}{c}{\multirow{-2}{*}{\textbf{Transformation}}} & \multicolumn{1}{r}{\textbf{P@1}} & \multicolumn{1}{r}{\textbf{P@3}} & \multicolumn{1}{r}{\textbf{P@50}} \\ \hline
None                                                        & None                                                          & \cellcolor[HTML]{FFFFFF}0.60        & \cellcolor[HTML]{FFFFFF}0.77        & \cellcolor[HTML]{FFFFFF}0.88         \\
Compiler                                                    & None                                                          & \cellcolor[HTML]{EDF5E8}0.70        & \cellcolor[HTML]{F1F8ED}0.83        & \cellcolor[HTML]{EBF4E6}0.90         \\
Compiler                                                    & Fixed Example                                                 & \cellcolor[HTML]{E2EFDA}0.76        & \cellcolor[HTML]{63BE7B}0.89        & \cellcolor[HTML]{83CB93}0.93         \\
Compiler                                                    & Smart Selection                                               & \cellcolor[HTML]{87CC96}0.81        & \cellcolor[HTML]{E2EFDA}0.88        & \cellcolor[HTML]{C3E3C3}0.92         \\
Abstracted Compiler                                         & None                                                          & \cellcolor[HTML]{F5FAF2}0.66        & \cellcolor[HTML]{EDF5E8}0.84        & \cellcolor[HTML]{E5F1DE}0.91         \\
Abstracted Compiler                                         & Fixed Example                                                 & \cellcolor[HTML]{DBEDD5}0.76        & \cellcolor[HTML]{63BE7B}0.89        & \cellcolor[HTML]{63BE7B}0.93         \\
Abstracted Compiler                                         & Smart Selection                                               & \cellcolor[HTML]{63BE7B}0.82        & \cellcolor[HTML]{63BE7B}0.89        & \cellcolor[HTML]{E2EFDA}0.92        
\end{tabular}
\caption{Excel: Results with different configurations.}
\label{table:config_excel}

\end{table}

\begin{table}[t!]
\small
\centering
\begin{tabular}{llrrr}
\hline
\multicolumn{1}{c}{}                                        & \multicolumn{1}{c}{}                                          & \multicolumn{3}{c}{\textbf{Metrics: Exact Match}}                                                       \\ \cline{3-5} 
\multicolumn{1}{c}{\multirow{-2}{*}{\textbf{Localization}}} & \multicolumn{1}{c}{\multirow{-2}{*}{\textbf{Transformation}}} & \multicolumn{1}{r}{\textbf{P@1}} & \multicolumn{1}{r}{\textbf{P@3}} & \multicolumn{1}{r}{\textbf{P@50}} \\ \hline
None                                                        & None                                                          & \cellcolor[HTML]{FFFFFF}0.47     & \cellcolor[HTML]{FFFFFF}0.68     & \cellcolor[HTML]{FFFFFF}0.84      \\
Compiler                                                    & None                                                          & \cellcolor[HTML]{F7FBF5}0.53     & \cellcolor[HTML]{F2F8EE}0.75     & \cellcolor[HTML]{E7F2E1}0.87      \\
Compiler                                                    & Fixed Example                                                 & \cellcolor[HTML]{88CD97}0.70     & \cellcolor[HTML]{E2EFDA}0.82     & \cellcolor[HTML]{63BE7B}0.89      \\
Compiler                                                    & Smart Selection                                               & \cellcolor[HTML]{63BE7B}0.71     & \cellcolor[HTML]{63BE7B}0.85     & \cellcolor[HTML]{E2EFDA}0.87      \\
Abstracted Compiler                                         & None                                                          & \cellcolor[HTML]{F8FBF6}0.53     & \cellcolor[HTML]{EEF6E9}0.77     & \cellcolor[HTML]{ECF5E7}0.86      \\
Abstracted Compiler                                         & Fixed Example                                                 & \cellcolor[HTML]{63BE7B}0.71     & \cellcolor[HTML]{63BE7B}0.85     & \cellcolor[HTML]{63BE7B}0.89      \\
Abstracted Compiler                                         & Smart Selection                                               & \cellcolor[HTML]{E2EFDA}0.68     & \cellcolor[HTML]{96D2A1}0.84     & \cellcolor[HTML]{E2EFDA}0.87     
\end{tabular}
\caption{PowerFx: Results with different configurations.}
\label{table:config_powerfx}

\end{table}

\begin{table}[t!]
\small
\centering
\begin{tabular}{llrrr}
\hline
\multicolumn{1}{c}{}                                        & \multicolumn{1}{c}{}                                          & \multicolumn{3}{c}{\textbf{Metric: Exact Match}}                                                        \\ \cline{3-5} 
\multicolumn{1}{c}{\multirow{-2}{*}{\textbf{Localization}}} & \multicolumn{1}{c}{\multirow{-2}{*}{\textbf{Transformation}}} & \multicolumn{1}{r}{\textbf{P@1}} & \multicolumn{1}{r}{\textbf{P@3}} & \multicolumn{1}{r}{\textbf{P@50}} \\ \hline
None                                                        & None                                                          & \cellcolor[HTML]{FFFFFF}0.19     & \cellcolor[HTML]{FFFFFF}0.28     & \cellcolor[HTML]{FFFFFF}0.39      \\
Compiler                                                    & None                                                          & \cellcolor[HTML]{EDF5E8}0.35     & \cellcolor[HTML]{ECF5E6}0.45     & \cellcolor[HTML]{EEF6E9}0.52      \\
Compiler                                                    & Fixed Example                                                 & \cellcolor[HTML]{BEE2BF}0.44     & \cellcolor[HTML]{B8DFBB}0.55     & \cellcolor[HTML]{E2EFDA}0.60      \\
Compiler                                                    & Smart Selection                                               & \cellcolor[HTML]{76C589}0.46     & \cellcolor[HTML]{63BE7B}0.59     & \cellcolor[HTML]{70C385}0.64      \\
Abstracted Compiler                                         & None                                                          & \cellcolor[HTML]{EAF3E4}0.37     & \cellcolor[HTML]{ECF5E7}0.45     & \cellcolor[HTML]{EEF6E9}0.52      \\
Abstracted Compiler                                         & Fixed Example                                                 & \cellcolor[HTML]{E2EFDA}0.43     & \cellcolor[HTML]{E2EFDA}0.53     & \cellcolor[HTML]{D6EBD1}0.60      \\
Abstracted Compiler                                         & Smart Selection                                               & \cellcolor[HTML]{63BE7B}0.47     & \cellcolor[HTML]{83CB93}0.58     & \cellcolor[HTML]{63BE7B}0.65     
\end{tabular}
\caption{Javascript: Results with different configurations.}
\label{table:config_js}
\end{table}

\begin{table}[t!]
\centering
\small
\begin{tabular}{llrrr}
\hline
\multicolumn{1}{c}{}                                        & \multicolumn{1}{c}{}                                          & \multicolumn{3}{c}{\textbf{Metric: Passes Parser}}                                                        \\ \cline{3-5} 
\multicolumn{1}{c}{\multirow{-2}{*}{\textbf{Localization}}} & \multicolumn{1}{c}{\multirow{-2}{*}{\textbf{Transformation}}} & \multicolumn{1}{r}{\textbf{P@1}} & \multicolumn{1}{r}{\textbf{P@3}} & \multicolumn{1}{r}{\textbf{P@50}} \\ \hline
None                                                        & None                                                          & \cellcolor[HTML]{FFFFFF}0.87     & \cellcolor[HTML]{F1F8ED}0.94     & \cellcolor[HTML]{63BE7B}0.98      \\
Compiler                                                    & None                                                          & \cellcolor[HTML]{F4F9F1}0.88     & \cellcolor[HTML]{FFFFFF}0.94     & \cellcolor[HTML]{E2EFDA}0.97      \\
Compiler                                                    & Fixed Example                                                 & \cellcolor[HTML]{C6E5C5}0.91     & \cellcolor[HTML]{C3E3C3}0.95     & \cellcolor[HTML]{ECF5E7}0.96      \\
Compiler                                                    & Smart Selection                                               & \cellcolor[HTML]{63BE7B}0.94     & \cellcolor[HTML]{63BE7B}0.97     & \cellcolor[HTML]{A3D7AB}0.97      \\
Abstracted Compiler                                         & None                                                          & \cellcolor[HTML]{E2EFDA}0.90     & \cellcolor[HTML]{E2EFDA}0.95     & \cellcolor[HTML]{E2EFDA}0.97      \\
Abstracted Compiler                                         & Fixed Example                                                 & \cellcolor[HTML]{D4EAD0}0.90     & \cellcolor[HTML]{E2EFDA}0.95     & \cellcolor[HTML]{ECF5E7}0.96      \\
Abstracted Compiler                                         & Smart Selection                                               & \cellcolor[HTML]{EEF6E9}0.89     & \cellcolor[HTML]{FFFFFF}0.94     & \cellcolor[HTML]{FFFFFF}0.95     
\end{tabular}
\caption{Python: Results with different configurations.}
\label{table:config_python}
\end{table}
\begin{table*}[t!]
	\centering
	\begin{tabular}{ccrrrc}
		\hline
		\multicolumn{1}{l}{\textbf{Language}} & \multicolumn{1}{l}{\textbf{Approach}}      &  \textbf{Pass@1} & \textbf{Pass@3} & \textbf{Pass@50} & \multicolumn{1}{l}{\textbf{Evaluation}} \\ \hline
		& RING (Compiler Message, Message Embedding) & 0.82                                         & 0.88                                         & 0.94                                          &                                         \\
		\multirow{-2}{*}{C}                   & Dr. Repair                                 & 0.63                                         & \multicolumn{1}{r}{-}                        & \multicolumn{1}{r}{-}                         & \multirow{-2}{*}{Passes Parser}         \\ \hline
		& RING (Compiler Message, Message Embedding) & 0.95                                         & 0.97                                         & 0.97                                          &                                         \\
		\multirow{-2}{*}{Python}              & BIFI                                       & 0.93                                         & 0.95                                         & 0.96                                          & \multirow{-2}{*}{Passes Parser}         \\ \hline
	\end{tabular}
	\caption{Results for two languages which do not have a ground truth -- C and Python, without token edit distance threshold.}
	\label{table:c_and_python_without_token_edit_dist}
\end{table*}

\begin{table}[t!]
\small
\centering
\begin{tabular}{llrrr}
\hline
\multicolumn{1}{c}{}                                        & \multicolumn{1}{c}{}                                          & \multicolumn{3}{c}{\textbf{Metric: Exact Matches}}                                                      \\ \cline{3-5} 
\multicolumn{1}{c}{\multirow{-2}{*}{\textbf{Localization}}} & \multicolumn{1}{c}{\multirow{-2}{*}{\textbf{Transformation}}} & \multicolumn{1}{r}{\textbf{P@1}} & \multicolumn{1}{r}{\textbf{P@3}} & \multicolumn{1}{r}{\textbf{P@50}} \\ \hline
None                                                        & None                                                          & \cellcolor[HTML]{F4F9F1}0.06     & \cellcolor[HTML]{FFFFFF}0.11     & \cellcolor[HTML]{FFFFFF}0.18      \\
Compiler                                                    & None                                                          & \cellcolor[HTML]{FFFFFF}0.04     & \cellcolor[HTML]{FDFEFC}0.12     & \cellcolor[HTML]{E7F2E0}0.21      \\
Compiler                                                    & Fixed Example                                                 & \cellcolor[HTML]{E2EFDA}0.09     & \cellcolor[HTML]{96D2A2}0.18     & \cellcolor[HTML]{E2EFDA}0.22      \\
Compiler                                                    & Smart Selection                                               & \cellcolor[HTML]{8ECF9B}0.10     & \cellcolor[HTML]{63BE7B}0.19     & \cellcolor[HTML]{63BE7B}0.27      \\
Abstracted Compiler                                         & None                                                          & \cellcolor[HTML]{F7FBF4}0.06     & \cellcolor[HTML]{FFFFFF}0.11     & \cellcolor[HTML]{FFFFFF}0.18      \\
Abstracted Compiler                                         & Fixed Example                                                 & \cellcolor[HTML]{63BE7B}0.11     & \cellcolor[HTML]{C9E6C7}0.17     & \cellcolor[HTML]{C9E6C7}0.23      \\
Abstracted Compiler                                         & Smart Selection                                               & \cellcolor[HTML]{E2EFDA}0.09     & \cellcolor[HTML]{E2EFDA}0.16     & \cellcolor[HTML]{E2EFDA}0.22     
\end{tabular}
\caption{Powershell: Results with different configurations.}
\label{table:config_powershell}
\end{table}

\begin{table}[t!]
\small
\centering
\begin{tabular}{llrrr}
\hline
\multicolumn{1}{c}{}                                        & \multicolumn{1}{c}{}                                          & \multicolumn{3}{c}{\textbf{Metric: Passes Parser}}                                                        \\ \cline{3-5} 
\multicolumn{1}{c}{\multirow{-2}{*}{\textbf{Localization}}} & \multicolumn{1}{c}{\multirow{-2}{*}{\textbf{Transformation}}} & \multicolumn{1}{r}{\textbf{P@1}} & \multicolumn{1}{r}{\textbf{P@3}} & \multicolumn{1}{r}{\textbf{P@50}} \\ \hline
None                                                        & None                                                          & \cellcolor[HTML]{FFFFFF}0.40     & \cellcolor[HTML]{FFFFFF}0.56     & \cellcolor[HTML]{FFFFFF}0.61      \\
Compiler                                                    & None                                                          & \cellcolor[HTML]{EBF4E5}0.52     & \cellcolor[HTML]{EBF4E5}0.63     & \cellcolor[HTML]{63BE7B}0.71      \\
Compiler                                                    & Fixed Example                                                 & \cellcolor[HTML]{E2EFDA}0.53     & \cellcolor[HTML]{F1F7ED}0.65     & \cellcolor[HTML]{F2F8EF}0.66      \\
Compiler                                                    & Smart Selection                                               & \cellcolor[HTML]{63BE7B}0.64     & \cellcolor[HTML]{63BE7B}0.69     & \cellcolor[HTML]{63BE7B}0.70      \\
Abstracted Compiler                                         & None                                                          & \cellcolor[HTML]{EBF4E5}0.53     & \cellcolor[HTML]{E2EFDA}0.64     & \cellcolor[HTML]{E4F0DC}0.68      \\
Abstracted Compiler                                         & Fixed Example                                                 & \cellcolor[HTML]{D3E9CF}0.56     & \cellcolor[HTML]{96D2A1}0.67     & \cellcolor[HTML]{E2EFDA}0.68      \\
Abstracted Compiler                                         & Smart Selection                                               & \cellcolor[HTML]{7BC88D}0.63     & \cellcolor[HTML]{63BE7B}0.68     & \cellcolor[HTML]{C3E3C3}0.69     
\end{tabular}
\caption{C: Results with different configurations.}
\label{table:config_c}
\end{table}

\end{document}